\documentclass[showpacs,preprintnumbers,amsmath,amssymb]{revtex4}
\usepackage{graphicx}
\usepackage{dcolumn}
\usepackage{amsmath}
\input{colordvi.tex}
\usepackage{color}

\newcommand{\simgt}{\lower.5ex\hbox{$\; \buildrel > \over \sim \;$}}
\newcommand{\simlt}{\lower.5ex\hbox{$\; \buildrel < \over \sim \;$}}
\newcommand{\bmf}[1]{\mbox{\boldmath$#1$}}

\def\apjs #1 #2 #3 {#1 ApJS, {\bf #2} #3}
\def\aap  #1 #2 #3 {#1 A\&A, {\bf #2} #3}
\def\mnras #1 #2 #3 {#1 MNRAS, {\bf #2} #3}
\def\physrep #1 #2 #3 {#1 Phys. Rep., {\bf #2} #3}

\def\h#1{\hbox{${}^{#1}$H}}

\def\h502{\hbox{$ h^{2}_{50}$}}

\def\fun#1#2{\lower3.6pt\vbox{\baselineskip0pt\lineskip.9pt
  \ialign{$\mathsurround=0pt#1\hfil##\hfil$\crcr#2\crcr\sim\crcr}}}
\begin{document}
%
\title{Constraints on Neutrino Masses from Weak Lensing}
\author{Kiyotomo Ichiki$^{1,2}$\footnote{E-mail address:
ichiki@a.phys.nagoya-u.ac.jp}, Masahiro Takada$^{3}$\footnote{E-mail address:
masahiro.takada@ipmu.jp}, Tomo Takahashi$^{4}$\footnote{E-mail address:
tomot@cc.saga-u.ac.jp}}
\affiliation{%
$^1$Department of Physics and Astrophysics, Nagoya University, Nagoya
464-8602, Japan
}
\affiliation{%
$^2$Research Center for the Early Universe, University of Tokyo, Tokyo 113-0033, Japan
}
\affiliation{%
$^3$Institute for the Physics and Mathematics of the Universe (IPMU), The University of Tokyo, Chiba 277-8582, Japan
}
\affiliation{%
$^4$Department of Physics, Saga University, Saga 840-8502, Japan
}
\date{\today}
\begin{abstract}
  The weak lensing (WL) distortions of distant galaxy images are
  sensitive to neutrino masses by probing the suppression effect on
  clustering strengths of total matter in large-scale structure. We
  use the latest measurement of WL correlations, the CFHTLS data, to
  explore constraints on neutrino masses.  We find that, while the WL
  data alone cannot place a stringent limit on neutrino masses due to
  parameter degeneracies, the constraint can be significantly improved
  when combined with other cosmological probes, the WMAP 5-year
  (WMAP5) data and the distance measurements of type-Ia supernovae
  (SNe) and baryon acoustic oscillations (BAO). The upper bounds on
  the sum of neutrino masses are $\sum m_\nu = 1.1$, 0.76 and 0.54 eV
  (95\% CL) for WL+WMAP5, WMAP5+SNe+BAO, and WL+WMAP5+SNe+BAO,
  respectively, assuming a flat $\Lambda$CDM model with finite-mass
  neutrinos.  In deriving these constraints, our analysis includes the
  non-Gaussian covariances of the WL correlation functions to properly
  take into account significant correlations between different angles.
\end{abstract}
\maketitle

\section{Introduction}
There are growing evidences that neutrinos have finite masses.  The
atmospheric, solar, reactor and accelerator neutrino oscillation
experiments have confirmed that at least two of the three neutrino
species have finite masses and their mass differences obtained from
current experiments are $|\Delta m^2_{21}|\approx 7.7\times
10^{-5}$~eV$^2$ and $|\Delta m^2_{31}|\approx 2.4\times
10^{-3}$~eV$^2$ \cite{2008NJPh...10k3011S}.  However, the absolute 
mass scale
has yet to be determined because these experiments are only
sensitive to the mass differences.  Whereas kinematical probes such as
tritium beta decay experiments 
and neutrinoless double beta decay experiments 
give upper bounds on the absolute neutrino 
mass 
\cite{Otten:2008zz,2002ARNPS..52..115E,2008arXiv0807.2457V},
cosmological observations provide a powerful, 
albeit indirect, means of constraining neutrino properties, indeed
yielding a more stringent upper bound on the total neutrino mass at
present.

The finite-mass neutrinos affect structure formation via their effects
on cosmic expansion history as well as on the evolution of
perturbations in structure formation, which imprint characteristic
signatures onto the cosmic microwave background (CMB) anisotropies and
the mass clustering history in large-scale structure at low redshifts
(see \cite{2006PhR...429..307L} for a thorough review). In particular
the neutrinos slow down the growth of total matter clustering compared
to a pure cold dark matter (CDM) model, because the neutrinos with
large thermal velocities cannot be trapped by the gravitational
potential well due to CDM inhomogeneities on small scales -- the
free-streaming effect \cite{1980PhRvL..45.1980B}. Combining the CMB
information with low-redshift cosmological probes allows to efficiently
track the suppression effect on gravitational clustering during the
matter-radiation equality and low redshifts
\cite{1998PhRvL..80.5255H,2006PhRvD..73h3520T}. In fact, with the
advent of high-precision cosmological probes, extensive efforts have
been made in order to derive a stringent upper limit on neutrino
masses by using/combining CMB and clustering properties of galaxies
and Lyman-$\alpha$ forests
\cite{2005PhRvD..71d3001I,2006PhRvD..74b7302F,2008arXiv0803.0547K,
  2006PhRvD..74l3507T,2006ApJ...647..799M,2007PhRvD..75h3510K,2008PhRvD..78h3535D,
  2006JCAP...10..014S,2008PhRvD..77h3507G}.

There is another vital probe of mass clustering: weak lensing (WL) or
the so-called cosmic shear, the bending of light by intervening mass
distribution that causes images of distant galaxies to be distorted
(see \cite{2001PhR...340..291B} for a thorough review). These sheared
source galaxies are mostly too weakly distorted  to measure the
effect on single galaxies, however, the lensing signals are measurable
by correlating the different galaxy images -- the shear correlation
functions. The weak lensing probes the distribution of total matter
along the line of sight, i.e. this method is free of galaxy bias
uncertainties. Various groups have measured the weak lensing
correlations and have also used 
the measurements
to constrain cosmological models since the first
detections
\cite{2000MNRAS.318..625B,2000astro.ph..3338K,2000Natur.405..143W,
  2000A&A...358...30V}. 
Moreover there are several studies that assess the
ability of future lensing surveys to constrain the neutrino masses
\cite{1999A&A...348...31C,2003PhRvL..91d1301A,2004PhRvD..70f3510S,2006JCAP...06..025H}.
However, the actual use of WL measurements for constraining the
neutrino masses has not been explored yet.

In this paper, therefore, we pursue a new constraint on neutrino
masses by using the latest WL data set, the Canada-France-Hawaii
Telescope Legacy Survey (CFHTLS) \cite{2008A&A...479....9F}, and also
by combining it with other cosmological probes, the WMAP 5-year data,
\cite{2008arXiv0803.0732H}, the type Ia supernova data
\cite{2008ApJ...686..749K}, and the baryon acoustic oscillation
experiment \cite{2005ApJ...633..560E}.  For this purpose, it is
critically important to take into account the covariances of the WL
correlation functions used, because there are significant
cross-correlations between the WL correlations of different angles --
the off-diagonal terms of the covariances are non-vanishing even for a
pure Gaussian field. We use the halo model approach developed in
\cite{TJ08} to compute the covariances, and also discuss the effect of
a possible uncertainty in the covariance estimation on the final
neutrino constraint.

The paper is organized as follows. In Section II-A we define the WL
correlation functions in terms of cosmological parameters and discuss
how finite-mass neutrinos affect the WL signals.  In Section II-B we
describe our model to compute the covariances of the WL correlation
functions.  After giving a brief description of other cosmological probes (CMB, SNe and
BAO) in Section II-D, we show the main results for the neutrino mass
constraints in Section III. Section IV is devoted to conclusions.

\section{Methodology}

\subsection{Preliminary}

In this section we briefly review how cosmic shear observable is
related to cosmology, with particular attention to the effect of
finite-mass neutrinos on cosmic shear.

The cosmic shear field is the weighted mass distribution integrated
along the line of sight.  The cosmic shear fields are measurable only
in a statistical sense.  A most conventional method used in the
literature is the two-point correlation based methods such as the
two-point correlations of the shear field.  The Fourier transformed
counterpart is the shear power spectrum, $P_\kappa$, which is obtained
by projecting the 3D matter power spectrum, $P_\delta(k)$, weighted
with the lensing kernel:
\begin{equation}
P_\kappa(\ell)=\frac{9 H_0^4 \Omega_m^2}{4c^2}\int_0^{\chi_{
 H}}\frac{d\chi}{a^2(\chi)}P_\delta\left(k=\frac{\ell}{f_K(\chi)};\chi\right)
\times \left[\int_\chi^{\chi_{H}}d\chi^\prime n(z)\frac{dz}{d\chi'}
\frac{f_K(\chi^\prime-\chi)}{f_K(\chi^\prime)}\right]^2
\label{eq:Pkappa},
\end{equation}
where $\chi$ is the comoving distance along the light ray, $\chi_{H}$
is the distance to the Hubble horizon, $\ell$ is the 2D wavevector
perpendicular to the line of sight, $H_0=100h~{\rm km~s}^{-1}~{\rm
  Mpc}^{-1}$ is the Hubble constant, and $f_K(\chi)$ is the comoving
angular diameter distance out to a distance $\chi$. Note that in
practice $\chi_H$ is taken to be the comoving distance out to a
redshift where a source galaxy distribution is sufficiently decaying.
The function $n(z)$ is the redshift distribution of source galaxies,
for which we adopt the following form characterized by three
parameters according to \cite{2008A&A...479....9F}:
\begin{equation}
n(z)=\frac{\beta}{z_s
 \Gamma \left(\frac{1+\alpha}{\beta}\right)}\left(\frac{z}{z_s}\right)^\alpha
 \exp\left[-\left(\frac{z}{z_s}\right)^\beta \right],
\label{eq:galaxy_dist}
\end{equation}
where $\Gamma(x)$ is the Gamma function. Note that $n(z)$ is
normalized so as to satisfy the condition $\int_0^\infty\!dz n(z)=1$.
The parameters $\alpha$, $\beta$, and $z_s$ are calibrated by using
the CFHT Deep Fields and the VIRMOS VLT Deep Survey
\cite{2006A&A...457..841I}.  Because the uncertainties in parameters
$\alpha$ and $\beta$ are small they were fixed to $0.838$ and $3.43$,
respectively, while the parameter $z_s$ will be marginalized over
assuming a Gaussian prior in the following analysis: $z_s=1.172\pm
0.026$, as was done in \cite{2007arXiv0712.1599D}.  For $l\simgt 100$
the major contribution to $P_\kappa(l)$ comes from nonlinear
clustering (e.g., see Fig.~2 in \cite{2004MNRAS.348..897T}).  As
described below, we employ the fitting formula in
\cite{2003MNRAS.341.1311S} to compute the nonlinear $P_\delta(k)$ from
the input linear matter power spectrum.  For the calculation of the
linear matter spectrum, we use the CAMB code \cite{Lewis:1999bs} up to
the wavenumber $k\leq k_{\rm split}=25$ Mpc$^{-1}$, and small scale
solutions of Hu \& Eisenstein \cite{1998ApJ...498..497H} for 
$k\geq k_{\rm split}$.  We chose this wavenumber because the bulk of
information in WL comes from wavenumbers up to $k/h \sim 1$
Mpc$^{-1}$ \cite{2002PhRvD..65f3001H}. We confirmed that the results
are stable against the change of $k_{\rm split}$.

The formulae to map the linear matter power spectra to the nonlinear
ones have been derived and tested against $N$-body simulations for CDM
cosmologies.  For a mixed dark matter model, the nonlinear matter
power spectrum has yet to be understood, except for the initial
attempts to study the weakly nonlinear regime of gravitational
clustering based on the perturbation theory \cite{2008PhRvL.100s1301S,2008JCAP...10..035W}
and the hybrid $N$-body simulation \cite{2008JCAP...08..020B}.  In
this paper we employ a similar prescription to that developed in
\cite{2008PhRvL.100s1301S} (also see \cite{2006JCAP...06..025H}) in
order to compute the nonlinear power spectrum for a mixed dark matter
model. Our method is as follows. First, recall that, for a range of
neutrino masses of interest, the finite mass neutrinos have large
free-streaming scale below which the density perturbation of neutrinos
is negligible; $k_{\rm fs}\sim 0.04(m_{\nu}/0.1~{\rm eV})(\Omega_{\rm
  m0}/0.27)^{1/2}~ h$Mpc$^{-1}$ (see Appendix A in
\cite{2006PhRvD..73h3520T} for the details), where $\Omega_{\rm m0}$
is the present-day energy density of total matter in units of the
critical density.  Hence we assume that the neutrino density
perturbations stay in the linear regime, which can be precisely
computed by solving the hierarchical Boltzmann equations coupled to
the Einstein equations using the CAMB. On the other hands, the
nonlinear clustering is driven mainly by the density perturbations of
CDM plus baryon. In this setting, the density perturbation of total
matter (CDM, baryon and neutrinos), which generates the gravitational
potential causing gravitational lensing via the Poisson equation, is
given as
\begin{equation}
\delta_{\rm tot}=\frac{\delta\rho_{\rm c}+\delta\rho_{\rm
 b}+\delta\rho_{\nu}}{\bar{\rho}_{\rm c}+\bar{\rho}_{\rm b}+\bar{\rho}_\nu}
=f_{\rm cb}\delta_{\rm cb}+f_\nu\delta^{\rm L}_\nu,
\label{eqn:delta_m}
\end{equation}
where 
\begin{equation}
 f_{\rm cb}\equiv \frac{\Omega_{\rm c0}+\Omega_{\rm b0}}{\Omega_{\rm
  m0}}~,~ f_\nu \equiv \frac{\Omega_{\nu0}}{\Omega_{\rm m0}}~,~
\delta_{\rm cb}\equiv\frac{\delta\rho_{\rm cb}+\delta\rho_{\rm
 b}}{\bar{\rho}_{\rm c}+\bar{\rho}_{\rm b}}~.
\end{equation}
The subscripts `c', `b', `$\nu$' and 'cb' stand for CDM, baryon,
finite mass neutrinos, and CDM plus baryon, respectively,
$\Omega_{i0}$ ($i={\rm c, b}$ and so on) denotes the present-day
energy density of the $i$-th component, $f_i$ is its fractional
contribution to the total energy density, and $\delta^{\rm L}_\nu$
represents the linear density perturbation of neutrinos.  Note
that $f_{\rm cb}+f_\nu=1$.  We neglect the gravitational dragging
force on nonlinear density perturbations of baryon and CDM from
neutrinos.  As stated above, we use the mapping formula of
\cite{2003MNRAS.341.1311S} to compute the nonlinear density
perturbation of CDM plus baryon from the input linear one.  Then the
nonlinear power spectrum of total matter is given by
\begin{equation}
 P^{\rm NL}_\delta(k)=\left<\delta_{\rm tot}^2\right> 
 =f_\nu^2 \left<(\delta_\nu^{\rm L})^2\right>+f_{\rm cb}^2\left<
(\delta_{\rm cb}^{\rm NL})^2\right>
	     +2f_\nu f_{\rm cb}\left<\delta_\nu^{\rm L} 
\delta^{\rm L}_{\rm cb}\right>~,
\label{eq:PNL}
\end{equation}
where we have introduced the superscript `NL' to explicitly state the
quantities in the nonlinear regime.  The sensitivity of the nonlinear
power spectrum depends to neutrino masses arises from the growth rate
in each density perturbations as well as from the dependence of each
terms in Eqn.~(\ref{eq:PNL}) on $f_\nu$.

One may imagine a more crude approach that the nonlinear power
spectrum is mapped by inserting the input linear power spectrum of
total matter into the fitting formula treating the total matter as a
single fluid. We have checked that the result obtained in this
approach is not so largely different from our fiducial approach of
Eqn.~(\ref{eq:PNL}) as long as neutrino masses are small. Even so we
believe that the method above is more sensible in a sense that the
method includes the perturbation theory result
\cite{2008PhRvL.100s1301S} when the perturbations are in the weakly
nonlinear regime.

\begin{figure}
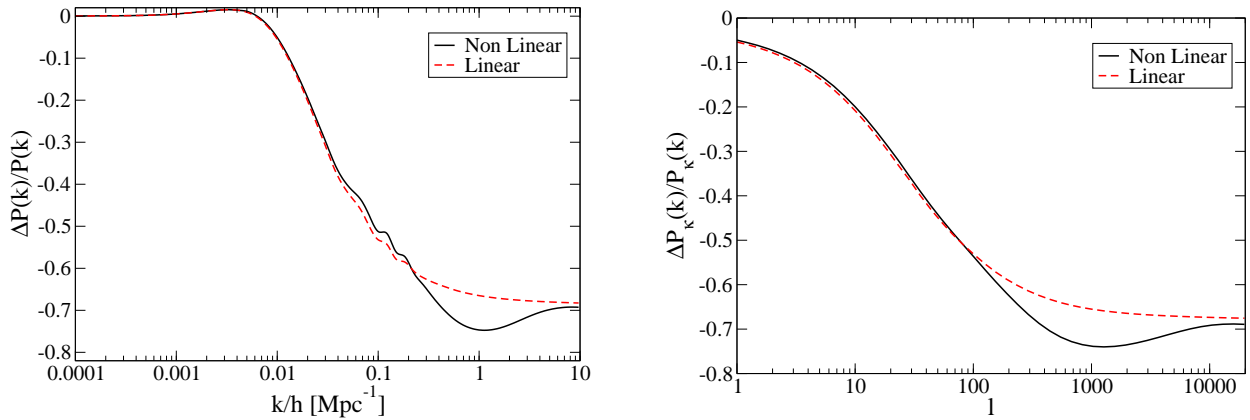

\begin{minipage}[m]{0.48\linewidth}
  \rotatebox{0}{\includegraphics[width=0.9\textwidth]{fig1-1.eps}}
\end{minipage}
\begin{minipage}[m]{0.48\linewidth}
  \rotatebox{0}{\includegraphics[width=0.9\textwidth]{fig1-2.eps}}
\end{minipage}
\vspace{0.4em}
\caption{{\em Left panel}: A fractional difference between the matter
  power spectra for a concordance $\Lambda$CDM model ($\Omega_{\rm
    m0}=0.3$) and a model with finite-mass neutrinos ($\Omega_{\rm
    cb0}=0.27$ and $\Omega_{\nu0}=0.03$). Note that the total matter
  density $\Omega_{\rm m0}(=\Omega_{\rm cb0}+\Omega_{\nu0})$ and other
  cosmological parameters are fixed for the two models. The solid
  curve shows the model prediction including the correction of
  nonlinear mass clustering (see the text for the details), while the
  dashed curve shows the linear-theory prediction. The finite-mass
  neutrinos cause a suppression in the power spectrum amplitudes on
  scales below the free-streaming scale.  The suppression effect is
  enhanced over transition scales between linear and nonlinear
  regimes.  {\em Right panel}: A similar plot, but for the lensing
  power spectrum as a function of multipoles $l$.}
 \label{fig:PkDiff}
\end{figure}
The effect of finite mass neutrinos on $P_\delta^{\rm NL}(k)$ is
depicted in the left panel of Fig.~\ref{fig:PkDiff}, showing the
relative difference between the nonlinear power spectra with and
without finite mass neutrinos, where $\Omega_{\rm m0}$ is kept
fixed. For comparison the corresponding result for the linear power
spectra is also shown, where the denominator and numerator in $\Delta
P_\delta/P_\delta$ are both the linear spectra. On very large distance
scales (i.e. very small $k$), well beyond the neutrino free-streaming
scale, the neutrino effect is absent, as the neutrinos can cluster
together with the CDM plus baryon perturbations.  For intermediate
scales, corresponding to $k\simeq [0.01-0.1]h$Mpc$^{-1}$ for the case
with $m_{\nu} \sim 0.5$~eV, the neutrinos cause a characteristic
scale-dependent suppression in the matter power spectrum amplitude.
Very interestingly, the neutrino suppression effect is enhanced on
scales around $k\sim 1h$Mpc$^{-1}$ compared to the linear theory
prediction, which is consistent with the perturbation theory result
found in \cite{2008PhRvL.100s1301S}.  Perhaps even more surprisingly,
the amount of the neutrino suppression effect becomes similar to that
of the linear theory prediction on the smaller scales $k\simgt
1h$Mpc$^{-1}$, well below the free-streaming scale. That is, the
neutrino effect appears just as a constant offset in the overall
amplitude on the small scales, even though the power spectrum
amplitude itself is significantly boosted in the nonlinear regime
compared with the linear theory. This seems consistent with the result
indicated by the hybrid $N$-body simulations
\cite{2008JCAP...08..020B}, although the simulation resolution may be
insufficient to follow this nonlinear regime. The amount of neutrino
suppression on scales down to nonlinear regime is roughly given by
$\Delta P_\delta/P_\delta\sim -8f_\nu$ which is sensitive to total
neutrino mass as $f_\nu\simeq \sum m_\nu/[\Omega_{\rm m0}
h^2(94.1~{\rm eV})]$, but not sensitive to the mass scale of
individual neutrinos.

There is a good rationale to believe that our modeling of the
nonlinear power spectrum is fairly reasonable according to the
previous findings in the literature.  As accounted for in the fitting
formula \cite{2003MNRAS.341.1311S}, the nonlinear spectrum amplitude
at a given scale is very sensitive to the local amplitude as well as
the local spectral slope $n_{\rm eff}=d\ln P_\delta^{\rm L}/d\ln k$ of
the input linear power spectrum at that scale, where the power
spectrum is a decreasing function with increasing $k$ at relevant
nonlinear scales (also see \cite{1996MNRAS.280L..19P}). In particular
a change in the local spectral slope induces a scale-dependent
modification in the shape of the nonlinear power spectrum 
over a range of the transition scales
between the linear and nonlinear regimes
(also
see Fig.~1 in \cite{1996MNRAS.280L..19P}).  Having in mind these
properties, the result in Fig.~\ref{fig:PkDiff} for the input linear
power spectra explicitly shows that including finite-mass neutrinos
leads to a change in the local spectral slope over a range of the
intermediate scales, causing the enhanced neutrino effect in the
weakly nonlinear regime. On very small scales, the local spectral
slope is unchanged, therefore, the neutrino effect becomes
scale-independent.

Thus the finite mass neutrinos affect the lensing power spectrum via
the effect on the matter power spectrum, even for a fixed $\Omega_{\rm
  m0}$, which keeps the lensing efficiency function unchanged. The
right panel of Fig.~\ref{fig:PkDiff} shows the neutrino effect on the
lensing spectra.  The neutrinos suppress the lensing spectrum
amplitudes over all the angular scales probed by the current data.
Note that the baryon oscillation wiggles are completely washed out due
to the projection.  Thus, since the shear correlation function is a
Fourier-transform of the lensing power spectrum, the neutrino effect
can be potentially extracted from the measured shear correlations. In
particular, if we can use the small angular scales with a robust model
prediction, the accuracy of constraining neutrino masses can be
significantly improved in the nonlinear regime compared to the linear
regime, due to the increased signal-to-noise ratio by the boosted
amplitudes of shear correlations.

\subsection{The CFHTLS result for cosmic shear measurement}

The data we will use to put limits on neutrino masses is taken from
the latest Canada-France-Hawaii Telescope Legacy Survey (CFHTLS or we
will often call WL for simplicity) \cite{2008A&A...479....9F}. It
covers over 57 square degrees, which drops to the effective area of
$34.2$ square degrees after taking into account the masking
corrections of the surveyed region, and comprises about $1.7\times
10^6$ galaxies whose shapes were used for the lensing shear correlation 
analysis. 
 In their paper, the shear correlation is given in terms of
correlation function $\xi$, $\gamma$, and $M_{\rm ap}$, all of which
are related to the same convergence power spectrum
(Eqn.~[\ref{eq:Pkappa}]) with different filtering.  They found that
these three statistics give consistent results on cosmological
parameter estimations. In this paper we therefore use one of the
correlation functions, $\xi$.

It is useful to decompose the cosmic shear correlations into the $E$
(irrotational) and $B$ (rotational) mode contributions, because the
cosmological shear field creates the $E$-mode field alone. The
$E$-mode correlation $\xi_E$ is related to the convergence power
spectrum $P_\kappa$
\cite{2002A&A...389..729S,2002ApJ...567...31P,2002ApJ...568...20C} as
\begin{equation}
\xi_E(\theta)=\int^\infty_0 \frac{\ell d\ell }{2\pi}
 P_\kappa(\ell) J_{0}(\ell \theta), 
\end{equation}
where $\theta$ is the separation angle between galaxy pairs, and
$J_{0}$ is the zero-th order Bessel function. In practice, since the
shear power spectrum is difficult to measure due to the complex survey
geometry, the $E$-mode correlation function is estimated from the
measured correlation functions $\xi_{\pm}(\theta)$ as proposed in
\cite{2002A&A...389..729S}:
\begin{eqnarray}
\xi_{E}(\theta)&=& \frac{1}{2}\left[ \xi_+(\theta) + \xi_-(\theta) +
\int_\theta^\infty \frac{d\theta^\prime}{
\theta^\prime}\xi_-(\theta^\prime)\left(4-12\frac{\theta^2}
{\theta^{\prime2}} \right) \right],
\end{eqnarray}
where $\xi_{\pm}(\theta)$ is the two-point correlations of
ellipticities between paired galaxies separated by angle $\theta$; the
`$+$' component is the tangential or radial ellipticity component with
respect to the line connecting the two galaxies, while the component
`$\times$' is its 45 degree rotated component. Thus, the
transformation relation of $\xi_E$ from the measured
$\xi_{\pm}(\theta)$ includes an integration of $\xi_-$ up to an
infinite separation $\theta\rightarrow \infty$.  However, in practice,
$\xi_E$ needs to be estimated from a finite range integration that may
introduce a residual uncertainty in the estimated $\xi_E(\theta)$. We
ignore this uncertainty in our paper for simplicity.

\begin{figure}
  \rotatebox{0}{\includegraphics[width=0.7\textwidth]{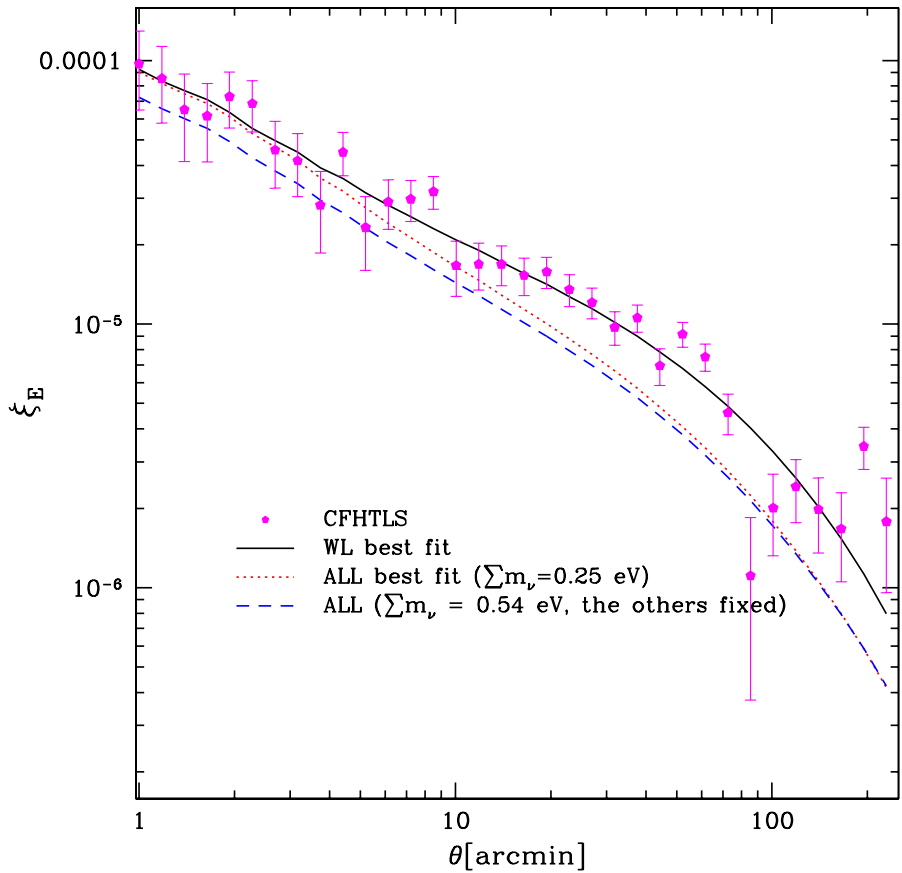}}
  \caption{The data points show the measured shear correlation
  functions, $\xi_E(\theta)$, at each angular bins, which are taken from
  the CFHTLS result in \cite{2008A&A...479....9F}. The error bars around
  each data points are computed from diagonal terms in the inverse of
  the covariance matrix that includes contributions from the shot noise
  of intrinsic galaxy ellipticities and the Gaussian and non-Gaussian
  sample variances (see the text for the details). The solid curve is
  the model prediction for the $\Lambda$CDM model with finite-mass
  neutrinos which best matches the WL measurement. The dotted curve is the
  best-fitting model prediction for the joint fitting of WL+WMAP5+SNe+BAO as
  will be shown below. Note that the best-fitting model has the total
  neutrino mass of $\sum\!m_\nu=0.25$~eV. To demonstrate the effect of
  finite-mass neutrinos on $\xi_E$, the dashed curve shows the model
  prediction where the neutrino mass is changed to $\sum
  m_\nu=0.54~$eV, roughly at two sigma upper bound for the joint
  fitting, and other cosmological parameters are fixed to their best-fitting
  values.  } \label{fig:xiEfig}
\end{figure}
Fig.~\ref{fig:xiEfig} shows the measured shear correlation function
$\xi_E$ taken from Table~B.1 in \cite{2008A&A...479....9F}, comparing
with the model predictions which best fit to to the CFHT WL data and
the case combined with the WMAP5 and other distance measurements (see
below for the details).  The error bars at each angular bins are
computed from the diagonal components of the inverse of 
the covariance matrix, and
the data points between different separation angles are significantly
correlated to each other as described below in detail.

\subsection{Covariances of shear correlation functions}
\label{sec:cov}

\begin{figure}
  \rotatebox{0}{\includegraphics[width=1.\textwidth]{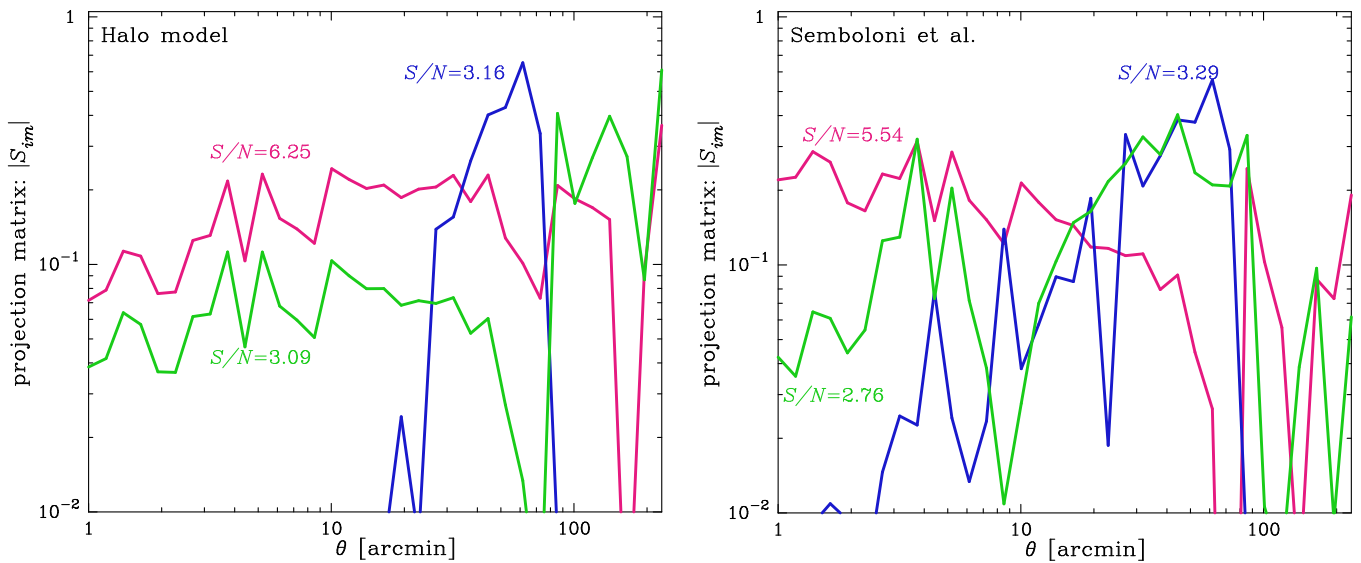}}
  \caption{The projection matrix $|S_{im}|$ for the principal
    component decomposition of the normalized covariance matrix is
    plotted, for first three eigenmodes that have largest differential
    contributions to the cumulative signal-to-noise ratio, $(S/N)$,
    defined by Eqn.(\ref{eqn:sn}). The left panel shows the results
    obtained when using the halo model developed in \cite{TJ08} to
    compute the covariance matrix. The total signal-to-noise ratio
    $S/N=10.6$.  The right panel shows the result when the fitting
    formula in Semboloni et al. (2007) \cite{2007MNRAS.375L...6S} is
    employed, yielding $(S/N)=8.89$.  For the real-space correlation
    function, the projection matrix for each eigenmodes has a broad
    tail, reflecting significant correlations between different
    angles.  It is also found that the contributions from each
    separation angles become different depending on which model of the
    covariance matrix is used.  } \label{fig:sn}
\end{figure}
The covariance matrix of shear correlation function quantifies (1)
statistical uncertainties in measuring the shear correlation function
at each angular bin, and (2) how the correlation functions of
different angular bins are correlated with each other.  Following the
method in \cite{2008A&A...482....9E,TJ08} (more explicitly see
Eqn.~[29] in \cite{TJ08}), the covariance matrix of the shear
correlation function, $\bmf{C}$, can be expressed as
\begin{eqnarray}
[\bmf{C}]_{ij}&\equiv&
{\rm Cov}[\xi_E(\theta_i),\xi_E(\theta_j)]\nonumber\\
&=&\delta^K_{ij}{\rm Cov}_{\rm
SN}+\frac{1}{\pi A_{\rm s}}\int_0^\infty\! ldlJ_0(l\theta_i)
J_0(l\theta_j)P_\kappa(l)+\frac{1}{4\pi^2A_{\rm
s}}\int_0^\infty\!ldl\int_0^\infty\!l'dl' 
J_0(l\theta_i)J_0(l\theta_j)\bar{T}_\kappa(l,l'), 
\label{eqn:cov}
\end{eqnarray}
where $\delta^{K}_{ij}$ denotes the Kronecker-type delta function
defined so that $\delta^{K}_{ij}=1$ when $\theta_i=\theta_j$ within
the bin width, otherwise $0$, $A_{\rm s}$ is the survey area in units
of steradian and $\bar{T}_\kappa$ is the angle averaged trispectrum of
convergence.  Note that the survey area is set to $A_{\rm s}=34.2$
square degrees as described above, and we have ignored the boundary
and geometry effects of surveyed region for simplicity.

The first term on the r.h.s. of Eqn.~(\ref{eqn:cov}),
$\delta^K_{ij}{\rm Cov}_{\rm SN}$, denotes the the shot noise
contamination due to intrinsic galaxy ellipticities, contributing only
to the diagonal terms of the covariance matrix with
$\theta_i=\theta_j$.  The shot noise is determined by the rms
intrinsic ellipticities and the total number of galaxy pairs that are
available from the surveyed region in the separation angle of a given
bin width. The shot noise can be estimated directly from the data: for
example, since the coherent shear signals can be erased by randomly
rotating each galaxy images by arbitrary angles, the shot noise
contamination can be estimated from variations in the correlation
functions that are repeatedly measured after randomization of galaxy
image orientations.  Our analysis uses the shot noise at each bin
given in the column labeled as ``$\delta\xi_B$'' in Table B1 of
\cite{2008A&A...479....9F}.

The second and third terms in Eqn.~(\ref{eqn:cov}) denote the Gaussian
and non-Gaussian sample variances, respectively, arising due to the
imperfect sampling of shear correlations from a finite survey region.
Since the lensing correlations probe the mass distribution in
large-scale structure along the line of sight, modeling the sample
variances requires a knowledge about the power spectrum and
trispectrum of the 3D mass distribution, where non-Gaussian errors
arise once the mass distribution probed by lensing resides in the
nonlinear clustering regime. Thus the sample variances depend on the
mass clustering strengths, i.e. on the underlying cosmology. There are
several important features of these sample variances. First, contrary
to the power spectrum covariance, the Gaussian term is non-vanishing
for the off-diagonal components of the covariance when $\theta_i\ne
\theta_j$: there are always correlations between the shear
correlations of different angles.  Second, the non-Gaussian sample
variance contributes to both the diagonal and off-diagonal terms of
the covariance matrix. The non-Gaussian errors become more significant
on smaller angular scales where the lensing signals are more affected
by the nonlinear regime.

Thus taking into account the covariance is critically important in
order not to have too optimistic parameter estimation from the
measured correlation functions.  However, only a few previous works
have studied the covariances of cosmic shear correlations, based on
ray-tracing simulations \cite{2007MNRAS.375L...6S} and analytic
methods \cite{2002A&A...396....1S,2008A&A...482....9E,TJ08}. In
particular, the importance of non-Gaussian errors is not fully
understood yet. In this paper, to estimate how an uncertainty in the
covariance affects the results, we employ two models of the
non-Gaussian covariances:
\begin{itemize}
\item[(A)] Compute the covariance matrix using the dark matter halo
  approach developed in \cite{TJ08}.
\item[(B)] Compute the non-Gaussian error contribution to the
  covariance by multiplying the fitting formula derived in
  \cite{2007MNRAS.375L...6S} with each matrix elements of the Gaussian
  sample variance (the second term in Eqn.[\ref{eqn:cov}]).
\end{itemize}
For the first approach we also include sources of new non-Gaussian
errors that inevitably arise for a {\em finite} survey area
\cite{2006MNRAS.371.1188H}, although the new contribution seems not so
significant as studied in \cite{TJ08}. Model A is our fiducial model
to compute the covariance, but we will study how our results change
when using Model B for the covariance evaluation. To compute the
covariance we need to specify cosmological
model, and our covariance evaluations assume the concordance
$\Lambda$CDM model which is consistent with WMAP5.

It is useful to study how the two-point correlation functions of
different separation angles are correlated with each other. To do
this, we study the cumulative signal-to-noise ($S/N$) ratio for
measuring the shear correlation functions.  Using the normalized
covariance matrix defined as $\tilde{C}_{ij}\equiv
C_{ij}/[\xi_E(\theta_i)\xi_E(\theta_j)]$, the $S/N$ is defined as
\begin{eqnarray}
\left(\frac{S}{N}\right)^2=\sum_{i,j=1}^{N_b}
[\tilde{\bmf{C}}^{-1}]_{ij}=\sum_{m=1}^{N_b}\frac{1}{\lambda_m}
\left(
\sum_{i=1}^{N_b}
S_{im}\right)^{2},
\label{eqn:sn}
\end{eqnarray}
where $\tilde{\bmf{C}}^{-1}$ is the inverse matrix of
$\tilde{\bmf{C}}$ and $N_b$ is the number of separation bins ($N_b=34$
for CFHTLS as shown in Fig.~\ref{fig:xiEfig}). In the second equality
we have used the principle component decomposition given as
$\tilde{C}_{ij}=\sum_{m=1}^{N_b}S_{im}\lambda_m S_{jm}$, where
$\lambda_m$ is the $m$-th eigenmode and the projection matrix $S_{im}$
describes how the $i$-th covariance element contributes to the $m$-th
eigenmode. Note that the inverse matrix of $\tilde{\bmf{C}}$ is given
by
$[\tilde{\bmf{C}}^{-1}]_{ij}=\sum_{m=1}^{N_b}S_{im}(1/\lambda_m)S_{jm}$,
and the matrix $S_{im}$ satisfies the conditions $\bmf{S}=\bmf{S}^{T},
\sum_{m=1}^{N_b}S_{im}S_{jm}=\delta_{ij}$, and
$\sum_{m=1}^{N_b}(S_{im})^2=1$. Eqn.~(\ref{eqn:sn}) expresses the
$S/N$ as a sum of the contributions from independent eigenmodes.

Fig.~\ref{fig:sn} shows the projection matrix of first three
eigenmodes that have largest differential contributions to the
$(S/N)^2$ in Eqn.~(\ref{eqn:sn}), using either of the two models above
to compute the covariance matrix. It is worth noting that Model A
gives $S/N=10.6$, while Model B gives $S/N=8.89$, because Model B
predicts more significant non-Gaussian errors than Model A as studied
in \cite{TJ08}. For both the models, each projection matrix gives
contributions over a wide range of separation angles, reflecting
significant correlations between different angles.  Thus there are
fewer independent modes than the 34 bins of separation angles in
Fig.~\ref{fig:xiEfig}: for both these two cases, first 15 eigenmodes
give about 95\% contribution to the $(S/N)^2$.  Also note that the
angular scales mainly contributing to the total $S/N$ are different in
these two models: for Model B based on the method in Semboloni et
al. (2007), the small angular scales are more important in
constraining cosmology.

We will use the log likelihood analysis to estimate cosmological
parameter:
\begin{equation}
\ln{\cal L}^{\rm WL}=-\frac{1}{2}\sum_{i,j=1}^{N_b}
\left[\xi_E^{\rm obs}(\theta_i)-\xi_E^{\rm model}(\theta_i)
\right]
[\bmf{C}^{-1}]_{ij}
\left[\xi_E^{\rm obs}(\theta_j)-\xi_E^{\rm model}(\theta_j)
\right],
\end{equation}
where $\xi_E^{\rm obs}$ and $\xi_E^{\rm model}$ are the measured
correlation function and the model prediction, respectively, and
$\bmf{C}^{-1}$ is the inverse of the covariance matrix given as
$[\bmf{C}^{-1}]_{ij}=\xi_{E}(\theta_i) [\tilde{\bmf{C}}^{-1}]_{ij}
\xi_{E}(\theta_j)$.

\subsection{Likelihood analysis}

Our likelihood analysis includes a fairly broad range of cosmological
parameters that affect a $\Lambda$CDM cosmology with finite mass
neutrinos. We will work on 8 parameters given as
\begin{equation}
\vec{p}=
(\Omega_{\rm b0} h^2, \Omega_{\rm dm0} h^2, f_\nu, 
\theta_A, \tau,  \ln (10^{10} \Delta_{\rm R}^2), n_s, z_s)~,
\end{equation}
where $\theta_A$ is the so-called acoustic scale related to the
distance to the last scattering surface \cite{2002PhRvD..66f3007K},
$\Omega_{\rm dm0} h^2 \equiv (\Omega_{c0}+\Omega_{\nu 0})h^2$ stands
for the dark matter density today, and $\Delta^2_{\cal R}$ and $n_s$
are the amplitude and spectral index of the primordial curvature power
spectrum defined at $k=0.002$ Mpc$^{-1}$ \cite{2008arXiv0803.0547K},
respectively. The parameter space we explore in the following is:
$\Omega_{\rm b0} h^2=[0.005, 0.1]$, $\Omega_{\rm dm0}
h^2=[0.01,0.99]$, $f_\nu=[0,0.9]$, $\theta_A=[0.3,10]$,
$\tau=[0.01,0.8]$, $\ln 10^{10}\Delta_{\cal R}^2=[0.5,6.0]$,
$n_s=[0.5,1.5]$, $z_s=[0.9,1.4]$.  Throughout this paper we assume the
standard three flavor neutrinos that are completely degenerate in
mass.  Note that the parameter $z_s$ specifies the galaxy redshift
distribution that is needed to predict the shear correlation functions
as described below Eqn.~(\ref{eq:galaxy_dist}), and does not affect
other cosmological probes such as WMAP5, SNe and BAO.  Assuming flat
priors for the cosmological parameters except $z_s$, we employ the
MCMC method \cite{2002PhRvD..66j3511L} to explore cosmological
parameter estimations in the multi-dimensional parameter space given
cosmological observables. For each run four parallel chains were
computed and the convergence test was made based on the Gelman and
Rubin statistics (``$R-1$'' statistics \cite{Gelman92}). Our chains
have typically 40000 points and $R-1 \lesssim 0.05$.  In what follows
we mention other cosmological probes used in our likelihood analysis,
which adds complementary information to the cosmic shear data.

\subsubsection{CMB}

The CMB anisotropies are generated mainly until epochs before the last
scattering surface of the decoupling epoch, $z\sim 1089$.  Hence, if
neutrino species are massive enough as $m_\nu \gtrsim 0.5$ eV (or
$\sum m_\nu\gtrsim 1.5$ eV), they became non-relativistic before the
recombination epoch \cite{2005PhRvD..71d3001I}.  In this case, the
finite-mass neutrino causes a significant effect on the CMB spectrum
mainly in two ways.  First, for a fixed energy density of total matter
(i.e. $\Omega_{\rm m0} =\Omega_{\rm cb0}+\Omega_{\nu0}$ fixed), the
matter-radiation equality is delayed by the presence of finite-mass
neutrinos. Hence this makes the heights of acoustic peaks higher,
especially that of the first peak, in the CMB spectrum due to a
greater early integrated Sachs-Wolfe effect.  Second, the neutrinos
cause an additional decay of the gravitational potential on small
scales.  This effect boosts the amplitudes of acoustic oscillations at
higher multipoles through the dilation effect
due to
diminishing gravitational potential \cite{1996ApJ...467...10D}.
Through these effects, a precise measurement of the CMB spectrum can
be used to place a limit on neutrino masses. In this paper we use the
WMAP 5-year (WMAP5) data, and explore the likelihood which can be
evaluated by the code publicly available at LAMBDA web site
\footnote{http://lambda.gsfc.nasa.gov/}.

For the masses below $m_\nu \lesssim 0.5$ eV ($\sum m_\nu\lesssim
1.5$eV), the neutrinos affect the CMB spectrum mainly through the
effect on the angular diameter distance out to the last-scattering
surface.  In this case the effect is degenerate with other
cosmological parameters such as $\Omega_{\rm m0}$ and $h$
\cite{2005PhRvD..71d3001I}.  Thus other cosmological probes
complementary to CMB are needed to break the parameter degeneracies in
order to explore the small mass scales of neutrinos.
\begin{figure}
  \rotatebox{0}{\includegraphics[width=0.6\textwidth]{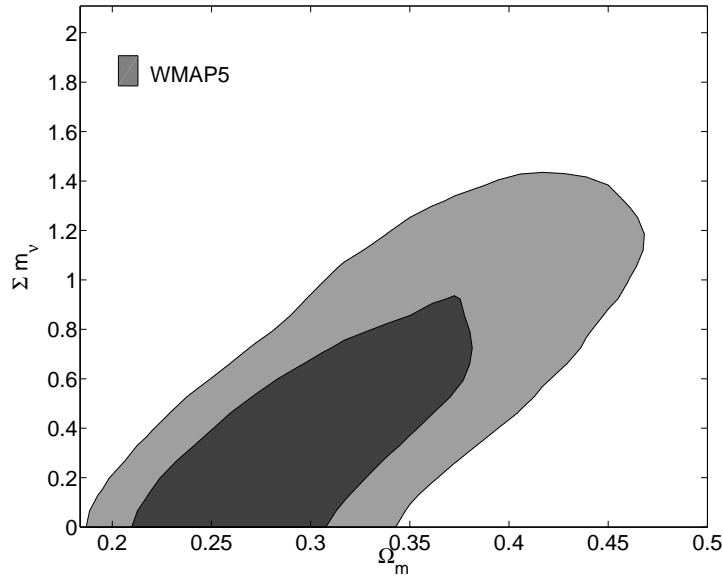}}
  \caption{The contours show the marginalized constraints (68\% and
    95\% CL) for the ($\sum m_\nu,\Omega_{\rm m0}$)-subspace, obtained
    by fitting the WMAP5 data to the $\Lambda$CDM model with
    finite-mass neutrinos. } \label{fig:Om_mnu}
\end{figure}

\subsubsection{Type Ia SNe}
The distant type Ia supernovae (SNe) are one of such probes. The use
of SNe as a standard candle allows us to estimate the luminosity
distance - redshift relation.  The estimated distance, more precisely
the distance modulus, can be compared with a model prediction for a
flat $\Lambda$CDM model with finite-mass neutrino contribution:
\begin{equation}
D_{\rm L}(z)=\frac{c(1+z)}{H_0}\int^z_0 dz^\prime 
\left[\Omega_{\rm r0} (1+z^\prime)^4+\Omega_{\rm cb0}(1+z^\prime)^3 +
 \Omega_\nu(z') + \Omega_{\Lambda0} \right]^{-1/2}~,
\label{eq:D_L}
\end{equation}
where $\Omega_{\rm r0}$ and $\Omega_{\Lambda0}$ are the energy
densities of photon and cosmological constant, respectively,
$\Omega_{\Lambda0}\approx 1-\Omega_{\rm m0}$ for a flat universe, and
$\Omega_\nu(z)$ is the energy density at a redshift $z$.  The
contribution from radiation is negligible for the luminosity distance
out to SNe at $z\lesssim 2$. In addition, since the current
cosmological probes are only sensitive to neutrino masses greater than
$\sim 0.1$eV as will be shown below, neutrinos in this mass scale
behave as non-relativistic matter over the redshift range of SNe data.
Therefore, if we restrict our attention to a flat $\Lambda$CDM
universe, the measured distance modulus of SNe constrain the
information mainly on the total matter density $\Omega_{\rm m0}$.  We
use the latest compilation of SNe data set given by
\cite{2008ApJ...686..749K}.  The SNe catalog provided consists of 307
SNe and is carefully calibrated taking into account various sources of
the systematic errors.

\subsubsection{BAO}
Baryon acoustic oscillation (BAO) offers another probe. CMB measures
the acoustic oscillations in the primordial photon-baryon plasma,
which can be used to derive the angular diameter distance to the last
scattering epoch, $z\approx 1089$.  The same oscillation feature is
imprinted onto the late-time matter power spectrum probed by a massive
galaxy survey \cite{1970ApJ...162..815P,1996ApJ...471..542H}.  This
baryon acoustic oscillation feature therefore provides a standard
ruler by which one can measure the distance to the redshift where the
bulk of galaxies are observed.  Eisenstein et
al. \cite{2005ApJ...633..560E} analyzed galaxies from Sloan Digital
Sky Survey (SDSS) and provided their result in a form of an effective
distance measure by averaging the distances in the radial and
transverse directions:
\begin{equation}
D_V(z)=\left[(1+z)^2 D_A^2(z) \frac{cz}{H(z)}\right]^{1/3}~,
\end{equation}
where $D_A(z)$ is the angular diameter distance to the redshift
$z=0.35$.  In this paper we use the following constraint on the
distance parameter $A$ provided in \cite{2005ApJ...633..560E}:
\begin{equation}
A(z=0.35)\equiv D_V\!(0.35)\frac{\sqrt{\Omega_{\rm m0} H_0^2}}{0.35c} =
 0.469\left(\frac{n_s}{0.98}\right)^{-0.35}\pm 0.017~.
\end{equation}
Notice that the value of $A$ is primarily sensitive to $\Omega_{\rm
  m0}$ for a flat universe case.

\section{Result and Discussion}

\subsection{Parameter constraints}
The WMAP team reported an upper limit on the total neutrino mass,
$\sum m_\nu<1.3$ eV (95\%CL), for a flat $\Lambda$CDM model plus one
additional parameter of the total neutrino mass
\cite{2008arXiv0803.0547K}.  Using the same likelihood function given
by WMAP team, we find a consistent but slightly tighter limit, $\sum
m_\nu<1.2$ eV.  This difference would be attributed to differences in
the treatment of cosmological parameters as well as the priors between
ours and WMAP5.  In fact, we have found the consistent result if the
same parametrization is adopted as theirs.  In any case, these upper
bounds are fairly close to the critical value $\sim 1.5$ eV as
discussed in the previous section.

Fig.~\ref{fig:Om_mnu} shows the results we obtained from the WMAP5
data: the inner and outer contours show the 68\% and 95\% confidence
level regions in the ($\Omega_{\rm m0}, \sum\! m_\nu$)-subspace,
marginalized over other parameters.  The contours show a strong
degeneracy between these two parameters, reflecting that the CMB
constraints come mainly from the effect of finite-mass neutrinos on
the matter-radiation equality.  The delay in the matter-radiation
equality caused by the presence of massive neutrinos is compensated by
adding more non-relativistic matter at present, i.e. increasing
$\Omega_{\rm m0}$.

The results in Fig.~\ref{fig:Om_mnu} are not so encouraging from the
perspective of complementarity between CMB and WL, because the WL
constraints on the parameters $\Omega_{\rm m0}$ and $\sum m_\nu$ are
expected to show a similar degeneracy curve.  The finite-mass
neutrinos cause suppression in the amplitudes of cosmic shear
correlations on relevant angular scales, however, this effect is
compensated by increasing $\Omega_{\rm m0}$ and/or the power spectrum
normalization (see Fig.~\ref{fig:PkDiff} and Eqn.~[\ref{eq:Pkappa}]).
Even so, since the degeneracy curves of CMB and WL constraints are not
elongated to exactly the same direction, combining these two can
always improve the constraints to some extent if they are consistent
with each other.  

\begin{figure}
  \rotatebox{0}{\includegraphics[width=0.7\textwidth]{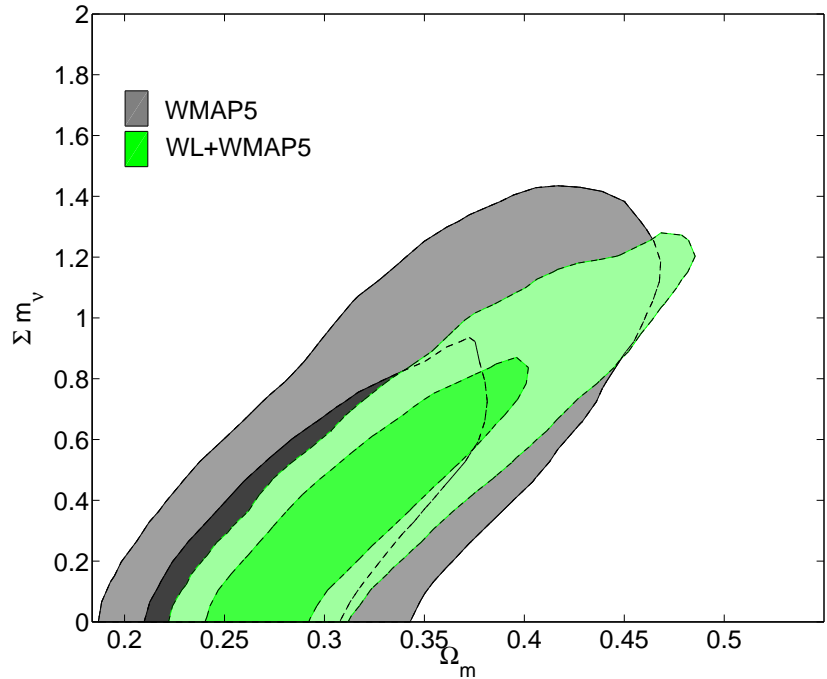}}
  \caption{The improvement in parameter constraints for ($\Omega_{m0}, \sum
    m_{\nu}$) obtained by combining the CFHT WL data with
    WMAP5 (green-color contours), in comparison with the constraints
    for WMAP5 alone (black).  The accuracy of $\Omega_{m0}$
    determination is improved by adding the WL constraint, because the
    WL amplitude is sensitive to $\Omega_{\rm m0}$ as can be found
    from Eqn.~(\ref{eq:Pkappa}). However, the constraint on $\Sigma
    m_\nu $ remains almost unchanged due to parameter
    degeneracies in the WL information.  }
\label{fig:WMAP_WL}
\end{figure}

Fig.~\ref{fig:WMAP_WL} shows the constraints on the
($\Omega_{\rm m0}, \sum m_\nu$)-plane obtained when combining WMAP5
with the CFHT WL constraints.  As expected from the above argument,
the constraint is only slightly improved when the WL data is included.
The upper bound on the total neutrino mass is $\sum m_\nu\simlt 1.1$
eV for WL+WMAP5 compared to $\sum m_\nu\simlt 1.2$ eV for WMAP5 alone.
The reason why the constraint is not significantly improved is also
attributed to the quality of WL data: the current WL data with $\sim
30$ square degrees is not so powerful yet compared to the WMAP5 data.
Finally, the weak lensing data itself suffers from a strong degeneracy
between $\Omega_{\rm m0}$ and the power spectrum normalization,
e.g. the $\Omega_{\rm m0}$-$\sigma_8$ degeneracy, as shown in
\cite{2008A&A...479....9F}: adding the WL data cannot yet much break
the degeneracy between $\sum m_\nu$ and $\Omega_{\rm m0}$ in the WMAP5
constraints.

Since the geometrical probes such as SNe and BAO are sensitive to
$\Omega_{\rm m0}$ for a flat $\Lambda$CDM model, by adding these
probes into the WMAP5 and WL data, the constraint on neutrino masses
can be further improved.  In fact the WMAP team nicely demonstrated
that combining SNe and BAO with the WMAP5 information tightens an
upper limit on the neutrino masses by a factor of 2, resulting in
$\sum m_\nu<0.67$ eV \cite{2008arXiv0803.0547K}.

\begin{figure}
  \rotatebox{0}{\includegraphics[width=0.8\textwidth]{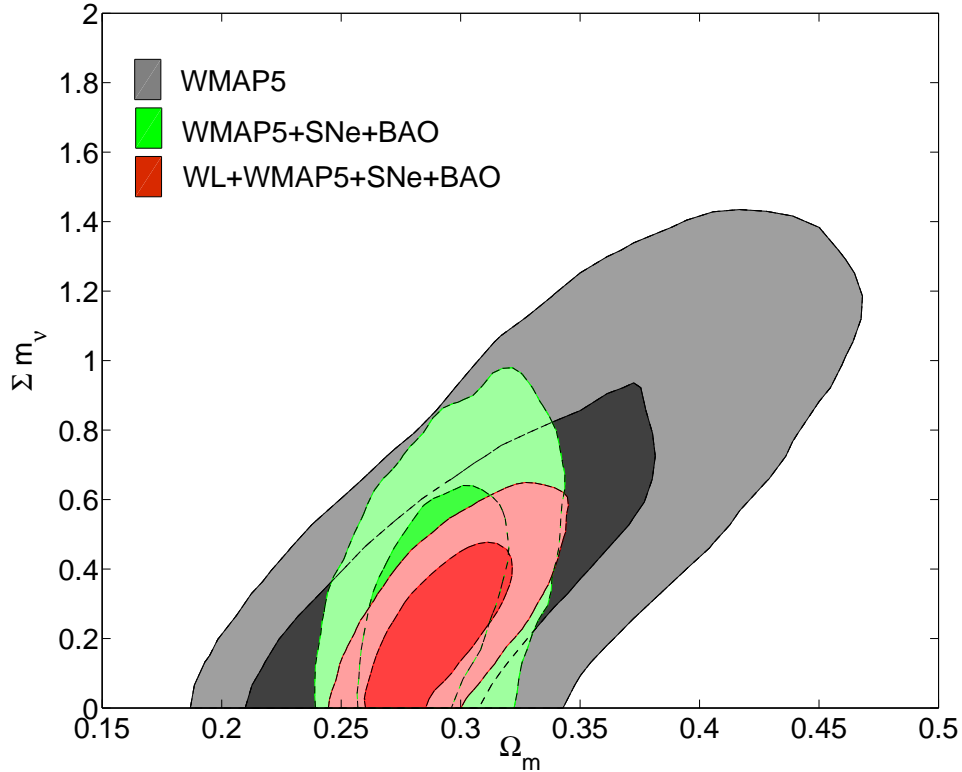}}
  \caption{Two-dimensional marginalized constraints on the neutrino
    mass $\sum m_\nu$ and $\Omega_{m0}$ are significantly improved
    when the constraints for WMAP5+WL are combined with the
    information from the geometrical probes, SNe
    and BAO. The black, green and red contours show the results for
    WMAP5, WMAP5+SNe+BAO, and WL+WMAP5+SNe+BAO, respectively. The
    marginalized upper bound on the neutrino mass is given as $\sum
    m_\nu\simlt 0.54$~eV for all the probes combined.
  }  \label{fig:Contour_all}
\end{figure}

Our analysis also confirmed this trend, but gives a slightly less
improved limit, $\sum m_\nu<0.76$ eV, as shown in
Fig.~\ref{fig:Contour_all}. The difference would be due to the fact
that we are using different BAO constraint from that used by the WMAP
team.  The BAO constraint employed here was one derived in Eisenstein
et al. \cite{2005ApJ...633..560E} using the spectroscopic sample of
the SDSS Luminous Red Galaxies (LRGs) that probes the distance out to
$z=0.35$.  This BAO measurement has a less constraining power than
that more recently derived by Percival et
al. \cite{2007MNRAS.381.1053P}, which was used in the WMAP5 analysis.
Since they used a bigger galaxy sample combining the SDSS LRGs with
the SDSS main galaxy sample and the 2dF galaxy sample, which probe the
distances to redshifts $z=0.2$ and $z=0.35$.  The constraints from
Percival et al. are more powerful than those from Eisenstein et al.;
for example, Percival et al. constrain $\Omega_{\rm m0}$ with the
precision of $\Omega_{\rm m0}=0.249\pm 0.018$ for a $\Lambda$CDM
model, while Eisenstein et al. give $\Omega_{\rm m0}=0.273\pm 0.025$.
In addition, as indicated from Fig.~\ref{fig:WMAP_WL}, the smaller
best-fitting value of $\Omega_{\rm m0}$ prefers a smaller $\sum m_\nu$
when the WMAP5 and BAO is combined.  However, a marginal tension
between the two data sets of BAO has been discussed in the literature,
which leads to a possible inconsistency for a $\Lambda$CDM model
(e.g., see Fig. 12 in \cite{2008arXiv0803.0547K}).  This tension is
not found if the BAO distance of $z=0.2$ is not included. For these
reasons, we employ the BAO constraint by Eisenstein et al. to derive a
conservative constraint on neutrino masses.

\begin{figure}
  \rotatebox{0}{\includegraphics[width=0.7\textwidth]{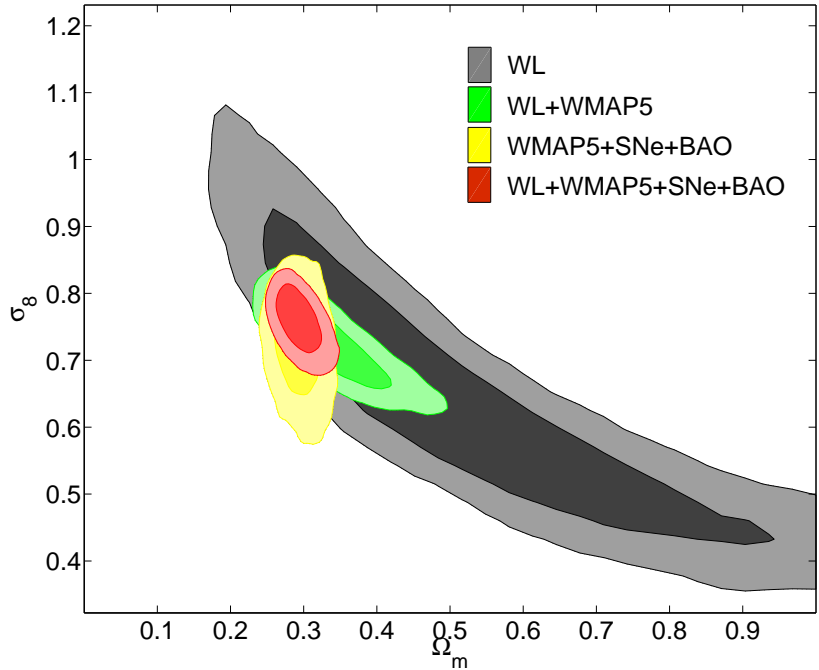}}
  \caption{Two-dimensional marginalized constraints on $\sigma_8$ and
    $\Omega_{m0}$, for WL+WMAP5 (green) and WL+WMAP5+SN+BAO (red),
    respectively, in comparison with the constraints from WL alone
    (black). The yellow contours show the region from WMAP5+SN+BAO
    (without WL).  } \label{fig:sig8-Om}
\end{figure}

Fig.~\ref{fig:Contour_all} shows how the neutrino mass constraints are
improved when some or all of WMAP5, WL, BAO and SNe constraints are
combined. Interestingly, adding the WL data into WMAP5+BAO+SNe does
improve the constraint on the total neutrino mass, leading to the
upper bound $\sum m_\nu<0.54$~eV, because the degeneracy with
$\Omega_{\rm m0}$ is more efficiently broken. It is also worth noting
that a most significant degeneracy inherent in the weak lensing
constraints, the $\Omega_{\rm m0}$--$\sigma_8$ degeneracy, is
efficiently broken by combining all the probes, as shown in
Fig.~\ref{fig:sig8-Om}. Our final results are $\Omega_{\rm
  m0}=0.294\pm 0.02$ and $\sigma_8=0.758\pm0.03$, improved by a factor
of 10 and 5 from the constraints by the WL data alone.

The shear correlation function for our best-fitting model is shown by the
dashed curve in Fig.~\ref{fig:xiEfig}, in comparison with the
measurement of WL and the model prediction (solid curve) that matches
the WL data alone. The prediction of the best-fitting model to all the data
combined (dotted curve) has systematically smaller amplitudes than the
measurement data points, especially on large angular scales. However, it
is worth noting that the change in $\chi^2$ between the two best-fitting
models for the joint fitting and the WL data alone is only $\Delta
\chi_{\rm WL}^2\simeq 2$ due to significant correlations between the
different data bins as implied in Fig.~\ref{fig:sn}. Adding more massive
neutrinos to the model with other parameters being fixed leads to
further smaller WL amplitudes compared to the data.  The dashed curve
shows the result for such a model assuming $\sum m_\nu=0.54$~eV, a mass
scale at about $2\sigma$ upper limit for the joint fitting, whose
$\chi^2$ value differs from the dotted curve by $\Delta\chi_{\rm
WL}^2\simeq 2$.

\begin{figure}
  \rotatebox{0}{\includegraphics[width=0.8\textwidth]{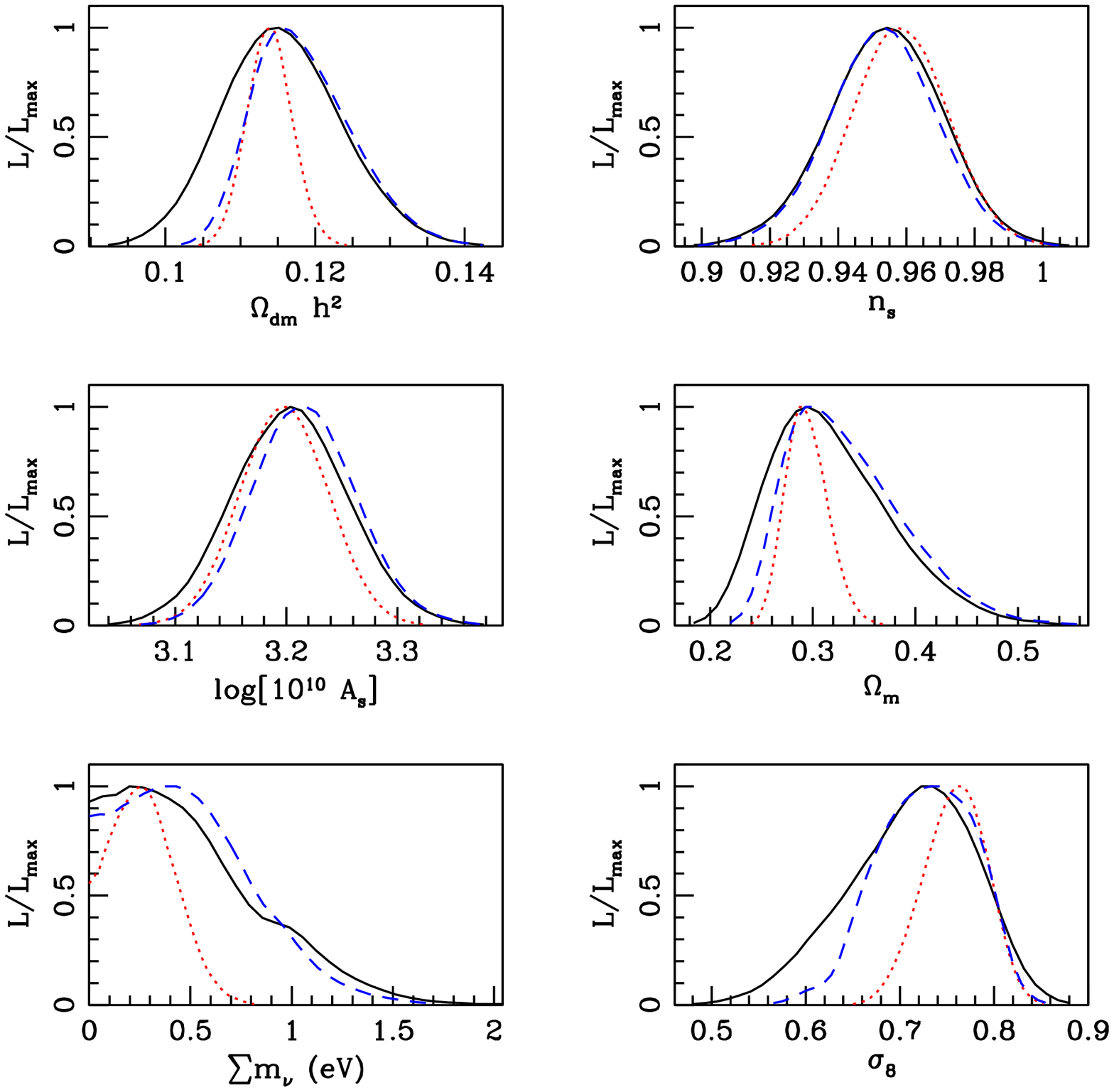}}
  \caption{One-dimensional marginalized likelihood on individual
    cosmological parameter for WMAP5 alone (solid curve), WL+WMAP5
 (blue dashed) and
    WL+WMAP5+BAO+SNe (red dotted), respectively. Some of parameters,
    $\Omega_{\rm m0}, \Omega_{\rm dm 0}h^2$, $\sigma_8$ and $\sum
    m_\nu$, are significantly improved by combining all the probes,
    where those parameters are sensitive to the overall amplitudes of
    cosmic shear correlations.  }
\label{fig:Compare_all}
\end{figure}

The marginalized constraints and likelihood distribution of individual
parameters of interest are summarized in Table~\ref{table:1} and
Fig.~\ref{fig:Compare_all}, respectively.  The precisions of
constraining some parameters such as $\Omega_{\rm m0}$, $\Omega_{\rm
  dm0}h^2$, $\sigma_8$ and $\sum m_\nu$ are significantly improved
when all the probes are combined.
\begin{table}[tbh]
\begin{center}
\caption{Summary of the constraints on cosmological parameters and the
 marginalized errors for $\nu\Lambda$CDM model}
{\begin{tabular}{c|@{\hspace{0.2in}}c@{\hspace{0.2in}}
|@{\hspace{0.2in}}c@{\hspace{0.2in}}|@{\hspace{0.2in}}c@{\hspace{0.2in}}
|@{\hspace{0.2in}}c@{\hspace{0.2in}}}
\toprule
Parameter & WL 
&WL+WMAP5 & WMAP5+SN+BAO & ALL  \\ \colrule
$\Omega_{\rm dm0}\, h^2 $&$0.22 \pm 0.09$&$0.12 \pm 0.0065$&$0.11 \pm 0.0038$&$0.11\pm 0.0030$ \\
$H_0$ [km s$^{-1}$ Mpc$^{-1}$]    &$74 \pm 15$    &$65 \pm 3.9$     &$68 \pm 2.0$     &$68\pm 1.9$\\
$n_s$               &$1.0\pm 0.26$  &$0.95 \pm 0.016$ &$0.96 \pm 0.014$ &$0.96\pm 0.014$\\
$\ln (10^{10}\Delta_{\cal R}^2)$   &$2.8\pm 1.0$   &$3.2 \pm 0.046$  &$3.2\pm 0.040$
	      &$3.2\pm 0.039$ \\
$\sum m_\nu$~[eV]       &$< 8.1$ (95\% CL)        &$< 1.1$          &$< 0.76$         &$< 0.54$ \\
$\sigma_8$          &$0.64\pm 0.15$ &$0.73 \pm 0.045$ &$0.73 \pm 0.059$
	      &$0.76 \pm 0.033$  \\
$\Omega_{\rm m0}$         &$0.52\pm 0.20$ &$0.33 \pm 0.056$ &$0.29 \pm 0.021$ &$0.29 \pm 0.020$  \\ \botrule
\end{tabular} \label{table:1}}
\end{center}
\end{table}

\subsection{Discussion of systematic errors}

There are several sources of systematic errors to affect weak lensing
measurements. One of those is uncertainties in the source galaxy
redshifts. As described in detail in \cite{2008A&A...479....9F}, the
redshift distribution of source galaxies is carefully calibrated by
using secure photometric redshifts of the CFHT deep fields, in
combination with the VIMORS spectroscopic sample, assuming that the
deep fields contain a representative galaxy sample of the WL galaxies.
We below estimate how a possible residual error in the source galaxy
redshifts, reported in \cite{2008A&A...479....9F}, affects our
neutrino constraint. Since the amplitude of cosmic shear correlations
is sensitive to mean redshift of galaxies, a most relevant parameter
in the source galaxy distribution is the parameter $z_s$ in
Eqn.~(\ref{eq:galaxy_dist}) for which we have so far assumed
$z_s=1.172 \pm 0.026$ for the fiducial value and the $1\sigma$
uncertainty.  Note that we have also employed the Gaussian prior
$\sigma(z_s)=0.026$ around the central value in our MCMC analysis.

\begin{figure}[h]
\begin{minipage}{0.48\textwidth}
  \rotatebox{0}{\includegraphics[width=0.9\textwidth]{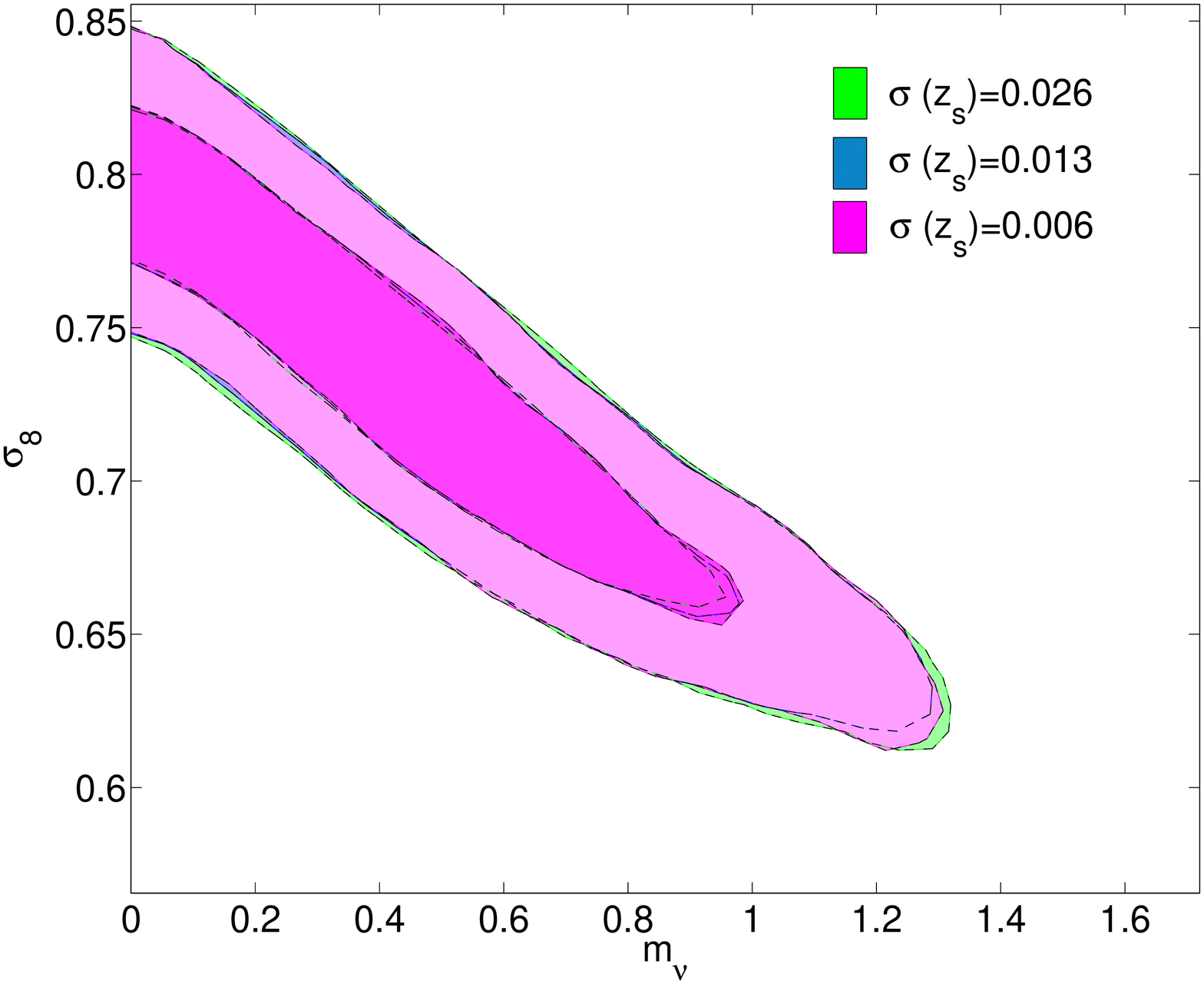}}
\end{minipage}
\begin{minipage}{0.48\textwidth}
  \rotatebox{0}{\includegraphics[width=0.9\textwidth]{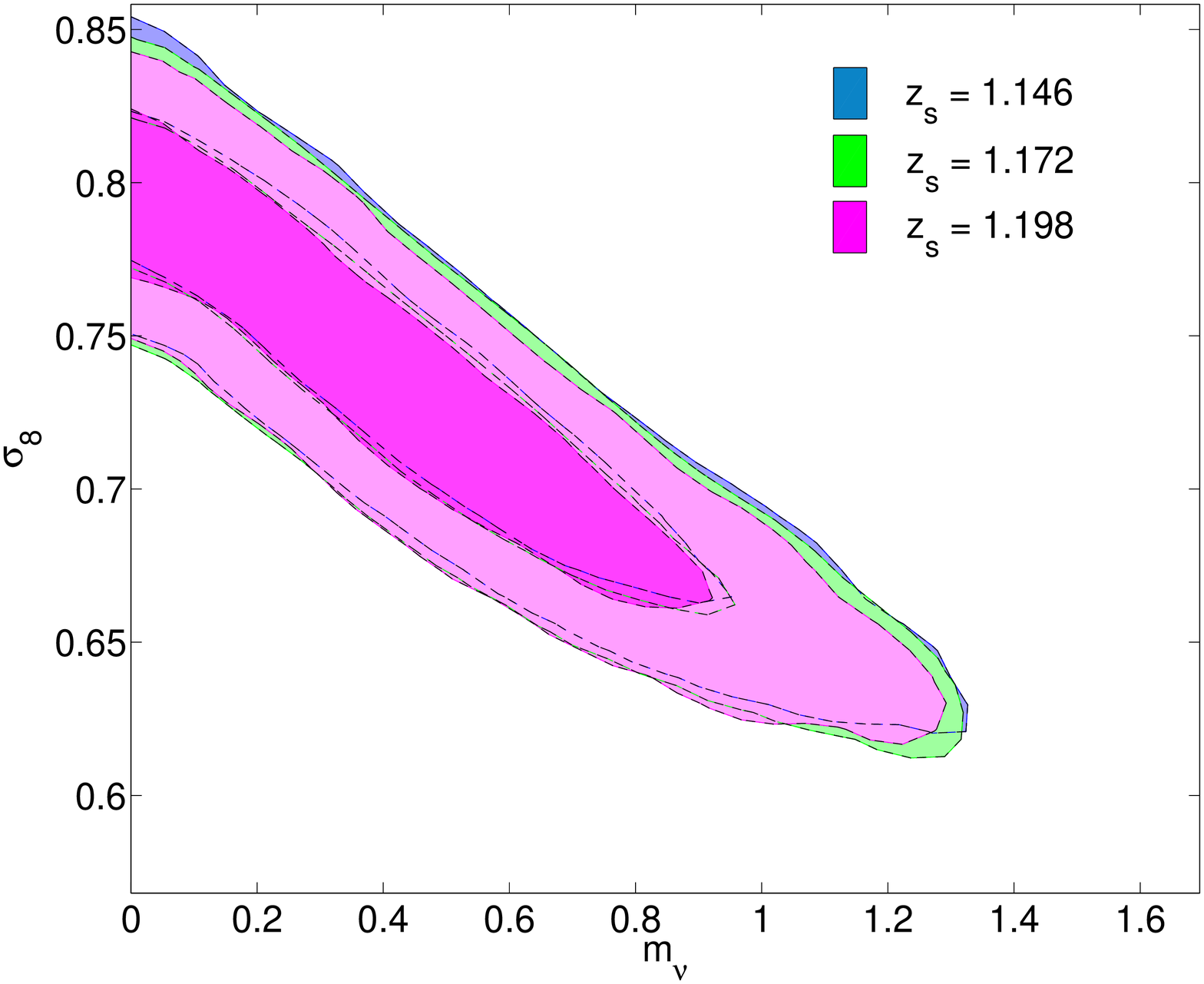}}
\end{minipage}
\caption{Effects of possible residual uncertainties in the mean
  redshift of weak lensing galaxies on the cosmological constraints
  on $(\sum m_\nu,\sigma_8)$ for WL+WMAP5. The mean redshift is parametrized by
  $z_s$ in Eqn.~(\ref{eq:galaxy_dist}), and the parameter is
  calibrated as $z_s=1.172\pm0.026$ according to
  \cite{2007arXiv0712.1599D}. We have so far assumed the Gaussian
  prior of $\sigma(z_s)=0.026$ around the central value in the
  parameter estimations.  {\em Left panel}: Shown is how the accuracy
  of the parameter estimation is changed if we assume a more restrict
  prior by a factor of 2 and 4 than the fiducial prior. The difference
  is indistinguishable, implying that this error is negligible
  compared to the measurement errors in shear correlations. {\em Right
    panel}: The effect of a shift in the central value of $z_s$ on the
  parameter constraints: the green and red contours show the results
  obtained when the central value of $z_s$ is changed by about
  $1\sigma$ statistical uncertainty to $z_s=1.146$ and $1.198$,
  respectively.  The difference is very small.  }
\label{fig:galaxy_dist}
\end{figure}

The left panel of Fig.~\ref{fig:galaxy_dist} shows how the constraints
on $\sum m_\nu$ and $\sigma_8$ change for the WL data combined with
WMAP5 if we put a more restrictive prior on $z_s$ by a factor of 2 or
4 than the fiducial prior. The effects appear to be small: the
confidence regions shrink only slightly.  This is simply because the
uncertainty in $z_s$ considered here is already sufficiently small
compared with the statistical uncertainties in the shear correlation
measurement.

The right panel of Fig.~\ref{fig:galaxy_dist} shows how a shift in the
central value of $z_s$ causes a bias in the best-fitting 
cosmological parameters, assuming the fiducial prior
$\sigma(z_s)=0.026$.  The smaller or larger source redshift gives larger
or smaller best-fitting values of $\sigma_8$, respectively (e.g., see
\cite{2003ApJ...597...98H,2008A&A...479....9F}), while the neutrino
constraint is little affected.  The effects of uncertainties in these
parameters on cosmological constraints are found to be very small ($\sim
1$\%) in the present analysis.  However, note that it will be of critical
importance to have a precise knowledge on source galaxy redshifts
especially when we have a sufficiently accurate measurement of cosmic
shear that is more powerful in constraining neutrino masses than the CMB
information \cite{2006JCAP...06..025H}.

\begin{figure}[h]
  \rotatebox{0}{\includegraphics[width=0.7\textwidth]{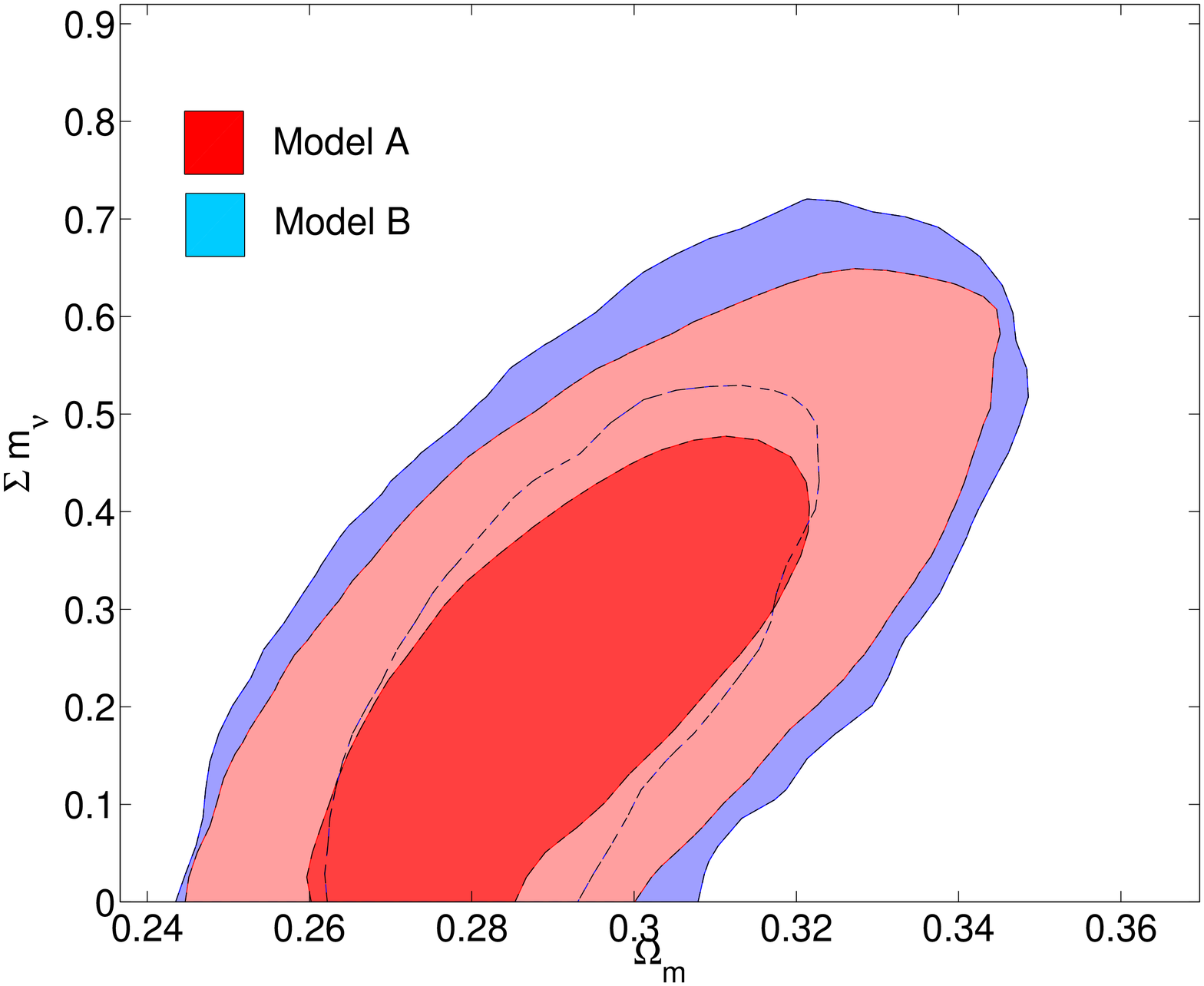}}
  \caption{Effects of possible uncertainties in non-Gaussian errors of
    the shear correlation covariances on the parameter constraints,
    for the ($\sum m_\nu, \Omega_{m0}$)-subspace.  Note that we
    consider the combined constraints of WL+WMAP5+SNe+BAO.  The red
    contours show the results when the halo model approach to compute
    the covariance (Model A: our fiducial model) is employed, while
    the blue contours show the results when the method developed in
    Semboloni et al. \cite{2007MNRAS.375L...6S} is employed (Model
    B). Model B predicts stronger non-Gaussian errors than Model A,
    resulting in larger confidence regions in these parameter space. The
    best-fitting parameters are almost identical between these
    two models.  } \label{fig:DifferentCovmat}
\end{figure}

As described in \S~\ref{sec:cov}, there is an uncertainty in
estimating the sampling variance contributions to the covariance
matrix of the shear correlation functions. We have so far used the
halo model approach developed in \cite{TJ08} to compute the covariance
matrix. Fig.~\ref{fig:DifferentCovmat} shows the constraints on $\sum
m_\nu$ and $\Omega_{\rm m0}$ obtained when using the different model
for the covariance matrix evaluation, Model~B based on
\cite{2007MNRAS.375L...6S}. These models are different in computing
the non-Gaussian error contributions to the covariance.  The
confidence regions for Model B are enlarged compared to those for
Model~A. The best-fitting parameters are only slightly changed: $\sum
m_\nu=0.25$ and $0.21$ for Model A and B, respectively.  Model B
predicts more significant non-Gaussian errors on angular scales of
$\theta \simlt 10$ arcminutes than Model A, i.e. stronger correlations
between the shear correlation functions of different angular bins. The
correlations reduce independent modes of the angular scales to be
useful for constraining cosmology. This explains larger confidence
regions for Model B. Also, as shown in Fig.~\ref{fig:sn}, different
angular scales of the shear correlation functions contribute to the
cosmological constraints in a different way between Model A and B,
causing a slight difference between the best-fitting parameters. 
Even so, it
can be found that, thanks to a wide coverage of angular scales for the
CFHT, the results for the two covariance models are consistent,
implying that the large angular scales where the Gaussian errors are
dominant in the covariance mainly contribute to the constraints on
cosmological parameters.

There are other sources of systematic errors such as those inherent in
the galaxy shape measurements. Estimating this effect is beyond the
scope of our paper, so we do not discuss  those effects here.

\section{Conclusion and Discussion}
In this paper we have investigated how the latest cosmic shear data
set, the CFHT weak lensing survey, can constrain neutrino masses. The
weak lensing correlations can probe clustering strengths of total
matter free of galaxy bias uncertainty, therefore, it allows us to
explore the suppression signal in the weak lensing amplitudes caused
by the free-streaming effect of finite-mass neutrinos. Since the
measured shear correlation functions do not have strong features in
the angular dependence, the shear amplitudes carry most of
cosmological information. However, due to the limited information for
the current data sets, the shear amplitudes integrated over a range of
source galaxy redshifts are only available: the redshift dependence of
the shear signals is not. For these reasons, constraining neutrino
masses from the weak lensing data alone suffers from severe
degeneracies among cosmological parameters such as $\sum m_\nu$,
$\Omega_{\rm m0}$ and the power spectrum normalization all of which
are sensitive to overall amplitudes of the shear signals. Yet,
interestingly enough, the neutrino mass constraint can be
significantly improved when the weak lensing constraint is combined
with other cosmological probes such as WMAP5, SNe and BAO, because the
combination of these data sets helps efficiently break parameter
degeneracies.  The upper limits on the total neutrino mass found in
this work are summarized as $\sum m_\nu<1.1$ eV (WL+WMAP5), $0.76$ eV
(WMAP5+SNe+BAO), $0.54$ eV (WL+WMAP5+SNe+BAO) at $95$\% confidence
level, respectively.  Note that we could not find any strong evidence
of the lower limit for neutrino masses, i.e. a detection of the
neutrino mass.  This might be attributed to insufficient statistical
precisions of the current data set ($\sim 30$ square degrees).

Throughout this paper we have assumed a minimal cosmological model: a
concordance $\Lambda$CDM model plus one additional parameter of the
total neutrino mass. Thus our approach derives the best-available
constraint on the neutrino mass from the weak lensing data combined
with other cosmological probes. The constraint will be relaxed if
additional parameters are included. In particular, dark energy
equation of state parameter ($w\equiv p_{\rm de}/\rho_{\rm de}$) may
be most intriguing to include, because a change in the equation of
state from the cosmological constant ($w=-1$) does change the cosmic
shear correlations in the shape and amplitude
\cite{2004MNRAS.348..897T}. The cosmological constraints from the
geometrical probes, SNe and BAO, are also relaxed if including $w$ as
a free parameter.  Therefore the accuracy of the neutrino mass
constraint is degraded if $w$ is included as a free parameter. This is
beyond the scope of this paper, and will be presented elsewhere.

Various weak lensing surveys are being planned and proposed so as to
have a much increased survey area by a factor of 100 compared to the
CFHT survey, corresponding to a survey with area more than a few
thousands square degrees. These surveys will promise to drastically
improve the statistical accuracies of the cosmic shear measurements by
a factor of 10 at each angular bins.  Our results of Table
\ref{table:1} imply that such a survey may allow a precision of
constraining the neutrino masses at level $\sigma(\sum m_\nu)\sim
0.8$~eV with the weak lensing data alone, if systematic errors
inherent in the measurements are well under control. Combining with
other probes allows to further improve the neutrino mass constraint as
demonstrated in this paper. In addition, if multi-color information is
available for future surveys as proposed, photometric redshift
information is available to estimate the redshift distribution of
source galaxies in a more reliable fashion.  This additional redshift
information is extremely useful in that it enables to recover redshift
dependence of the lensing signals, the so-called lensing tomography,
allowing to efficiently break parameter degeneracies such as
$\sigma_8$-$\Omega_{\rm m0}$ degeneracy with the weak lensing data
alone \cite{2004MNRAS.348..897T}. 
Therefore, if lensing tomography is
available, weak lensing (combined with CMB and others) would
potentially allow a detection of the neutrino masses.

In addition to the weak lensing technique, several cosmological probes 
have been proposed to constrain neutrino masses at accuracies of 
$O(0.1)$~eV, such as CMB
lensing 
\cite{2006PhRvD..73d5021L}, 21cm experiments \cite{2006ApJ...653..815M},
galaxy surveys \cite{2006PhRvD..73h3520T}, cluster number count
 \cite{2005PhRvL..95a1302W}, and so on.
All of these techniques are 
essentially based on the same effects of neutrinos, i.e., neutrinos
suppress the power of density fluctuations on small scales. Therefore, 
these experiments including weak lensing allow us to make cross-checks of
the cosmological constraint and hopefully to detect the finite mass of neutrinos, resolving the hierarchy of
neutrino masses  with cosmological data sets.

\bigskip When this paper was under completion, similar works for
constraining the neutrino mass with the CFHT data combined with other
cosmological probes were put forward by
\cite{2008arXiv0810.0555T,2008arXiv0810.3572G}. 
In \cite{2008arXiv0810.0555T}, they found an upper bound on
neutrino masses which is consistent with ours. However, they also
reported a lower bound $\sum m_\nu > 0.03$ eV which we could not find
here. The difference may be attributed to different data set used for BAO and
SNIa, and different covariance matrices from ours. In
\cite{2008arXiv0810.3572G} they derived a  
constraint on neutrino masses from CFHT data without combining WMAP5
data. Therefore their obtained upper bound is slightly weaker than ours
but it is a consistent result. After we submitted our paper, a similar paper
\cite{2008arXiv0812.1672L} was posted on the arXiv, in which further
data set such as SDSS LRG and the other CMB data were used to obtain the
upper bound on neutrino masses.

\acknowledgments We would like to thank K.~Ichikawa, B.~Jain,
A.~Kusaka, S.~Saito and A.~Taruya for useful discussion.  This work
was in part supported by World Premier International Research Center
Initiative (WPI Initiative), MEXT, Japan, by Grant-in-Aid for
Scientific Research on Priority Area No. 467 ``Probing Dark Energy
through an Extremely Wide and Deep Survey with Subaru Telescope'', by
Grant-in-Aid for Scientific Research Grant (Nos. 17740129, 18072001,
19740145 and 20740119), by the Sumitomo Foundation, by Grant-in-Aid
for the Global Center of Excellence program ``Quest for Fundamental
Principles in the Universe'' at Nagoya University, as well as by
Grant-in-Aid for the 21st Century Center of Excellence program
``Exploring New Science by Bridging Particle-Matter Hierarchy'' at
Tohoku University.


\bibliography{WL}
\end{document}